\documentclass[a4paper,twocolumn,10pt,accepted=2024-08-07]{quantumarticle}

\pdfoutput=1
\usepackage[utf8]{inputenc}
\usepackage[english]{babel}
\usepackage[T1]{fontenc}
\usepackage{amsmath}
\usepackage{hyperref}
\usepackage[numbers]{natbib}

\usepackage{bbold}
\usepackage{leftindex}

\newcommand{\tr}{\mbox{Tr}}

\newcommand{\ket}[1]{\left | #1 \right \rangle}
\newcommand{\ketNoResize}[1]{| #1 \rangle}
\newcommand{\bra}[1]{\left \langle #1 \right |}
\newcommand{\braL}[2]{\leftindex_{#1}{\bra{#2}}}

\newcommand{\beq}{\begin{equation}}
\newcommand{\eeq}{\end{equation}}

\newcommand{\ha}{{\hat\alpha}}
\newcommand{\hb}{{\hat\beta}}
\newcommand{\hc}{{\hat\gamma}}
\newcommand{\hd}{{\hat\delta}}
\newcommand{\hU}{{\hat U}}
\newcommand{\hV}{{\hat V}}
\newcommand{\hW}{{\hat W}}
\newcommand{\hR}{{\hat R}}
\newcommand{\hS}{{\hat S}}
\newcommand{\hT}{{\hat T}}
\newcommand{\hQ}{{\hat Q}}
\newcommand{\hSigma}{{\hat \sigma}}

\begin{document}

\title{Classical-to-quantum non-signalling boxes}
\author[1]{Carolina Moreira Ferrera}
\orcid{0009-0001-5605-0643}
\author[2]{Robin Simmons}
\orcid{0009-0003-5570-3512}
\author[3]{James Purcell}
\orcid{0000-0002-6956-431X}
\author[1]{Daniel Collins}
\orcid{0009-0009-0365-4206}
\author[1]{Sandu Popescu}
\orcid{0000-0003-1142-7694}
\affil[1]{H. H. Wills Physics Laboratory, University of
Bristol, Tyndall Avenue, Bristol BS8 1TL}
\affil[2]{Blackett Laboratory, Imperial College London, London SW7 2AZ, UK}
\affil[3]{Dept. of Physics $\&$ Astronomy, University College London, WC1E 6BT, UK}

\begin{abstract}
Here we introduce the concept of classical input - quantum output (C-Q) non-signalling boxes, a generalisation of the classical input - classical output (C-C) non-signalling boxes. We argue that studying such objects leads to a better understanding of the relation between quantum nonlocality and non-locality beyond quantum mechanics. The main issue discussed in the paper is whether there exist “genuine” C-Q boxes or all C-Q boxes can be built from objects already known, namely C-C boxes acting on pre-shared entangled quantum particles. We show that large classes of C-Q boxes are non-genuine. In particular, we show that all bi-partite C-Q boxes with outputs that are pure states are non-genuine. We also present various strategies for addressing the general problem, i.e. for multi-partite C-Q boxes which output mixed states, whose answer is still open.  Finally, we show that even some very simple non-genuine C-Q boxes require large amounts of C-C nonlocal correlations in order to simulate them. 

\end{abstract}

\maketitle

\section{Introduction}

During the recent years, the existence of long-distance nonlocal correlations, first discovered by J. Bell \cite{bell} and experimentally confirmed by S.J. Freedman and J.F. Clauser \cite{clauserExp} and A. Aspect, P.Grangier and G.Roger \cite{aspect} became to be understood as one of the main aspects of Nature. Very intensive research has taken place in the subject, from understanding the various aspects of entanglement and Bell inequalities to making use of nonlocality in virtually all of quantum information tasks and even to leading to new insights in quantum gravity.

To further compound the surprise of the very existence of nonlocality, it has been later realised that nonlocal correlations even stronger than those allowed by quantum mechanics could in principle exist without entering in conflict with relativity \cite{PR}. This has raised fundamental questions about Nature. Perhaps such correlations exist in Nature, only we have not discovered them yet. If discovered, it would mean that quantum mechanics is not a valid description of Nature and needs to be replaced by another theory. On the other hand, if such correlations do not exist, why don’t they exist? As they are not in contradiction with relativity, what other fundamental principles of Nature rule them out? 

One fruitful approach to the above question has been to consider “non-signalling boxes”, hypothetical boxes that accept classical inputs and yield classical outputs that are non-locally correlated with each other \cite{PRBarrett}. Then try to find tasks to which they would be useful when the correlations are stronger than those allowed by quantum mechanics. It has been discovered that some tasks, mostly of information processing nature, with no relation whatsoever with quantum mechanics, have qualitatively different behaviour when allowed access to such boxes \cite{vanDam,brassard,localOrthogonality,macroscopicLocality}. Tantalisingly, some tasks undergo a qualitative change precisely at the border of quantum to beyond quantum correlation strengths \cite{quantumLimitsFromSwapping, informationCausality, non-localComputation}. This already shows that quantum mechanics is a very special theory from a fundamental point of view, unrelated to “physical” properties such as structure of atoms, etc.  Yet not the entire boundary of the set of quantum correlations has been singled out in this way. It is therefore quite important, in order to make progress along this line, to find new tasks and/or different ways to characterise nonlocality. Many of these have been recently suggested, such as post-quantum steering \cite{ChannelFrameworkSteering,PostQuantumSteering,formalismSteering} and post-quantum GPTs \cite{GPTBarrett,hardy}. For extensive reviews see
\cite{resourceNonclassicalChannels,GPTReview}. 

\begin{figure}[ht]
\centering
\includegraphics[width=5cm]
{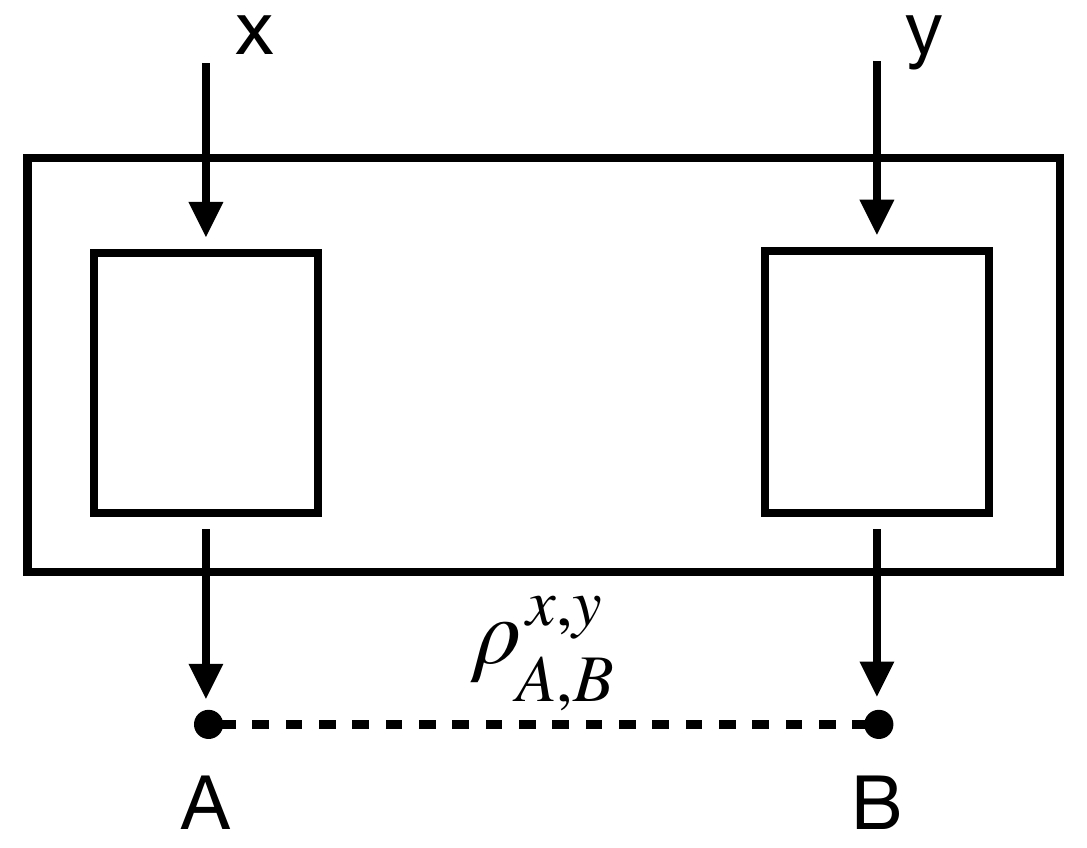}
\caption{A CQ box has two parts, taking inputs x and y.  It outputs a quantum state $\rho^{x,y}_{A,B}$, joint across A and B.}
\label{Fig:CQBoxes1}
\end{figure}

Here we introduce a (potentially) new type of non-signalling, non-local “boxes”. The boxes have classical inputs and they output quantum particles in given correlated states, depending on the inputs as shown in Fig \ref{Fig:CQBoxes1}. 

In order for the box to be non-signalling, the reduced density matrix of each party must be independent on the input of the other party: 
\beq \begin{split} 
\label{noSignallingConditions}
\rho^{x,y}_{A}=\rho^{x}_{A}\\ \rho^{x,y}_{B}=\rho^{y}_{B} 
\end{split} \eeq
where $\rho^{x,y}_{A}=\tr_B \rho^{x,y}_{A,B}$ and $\rho^{x,y}_{B}=\tr_A \rho^{x,y}_{A,B}$.  
In the case of multiple parties we generalise these conditions to say that the density matrices of all subgroups of parties could not depend on the inputs of the other parties. 

The present paper addresses what is arguably the most important question related to classical-quantum (C-Q) boxes: Do there exist {\it genuine} C-Q non-signalling boxes, or could all of them be decomposed in already known objects?  

\begin{figure}[ht]
\centering
\includegraphics[width=5cm]{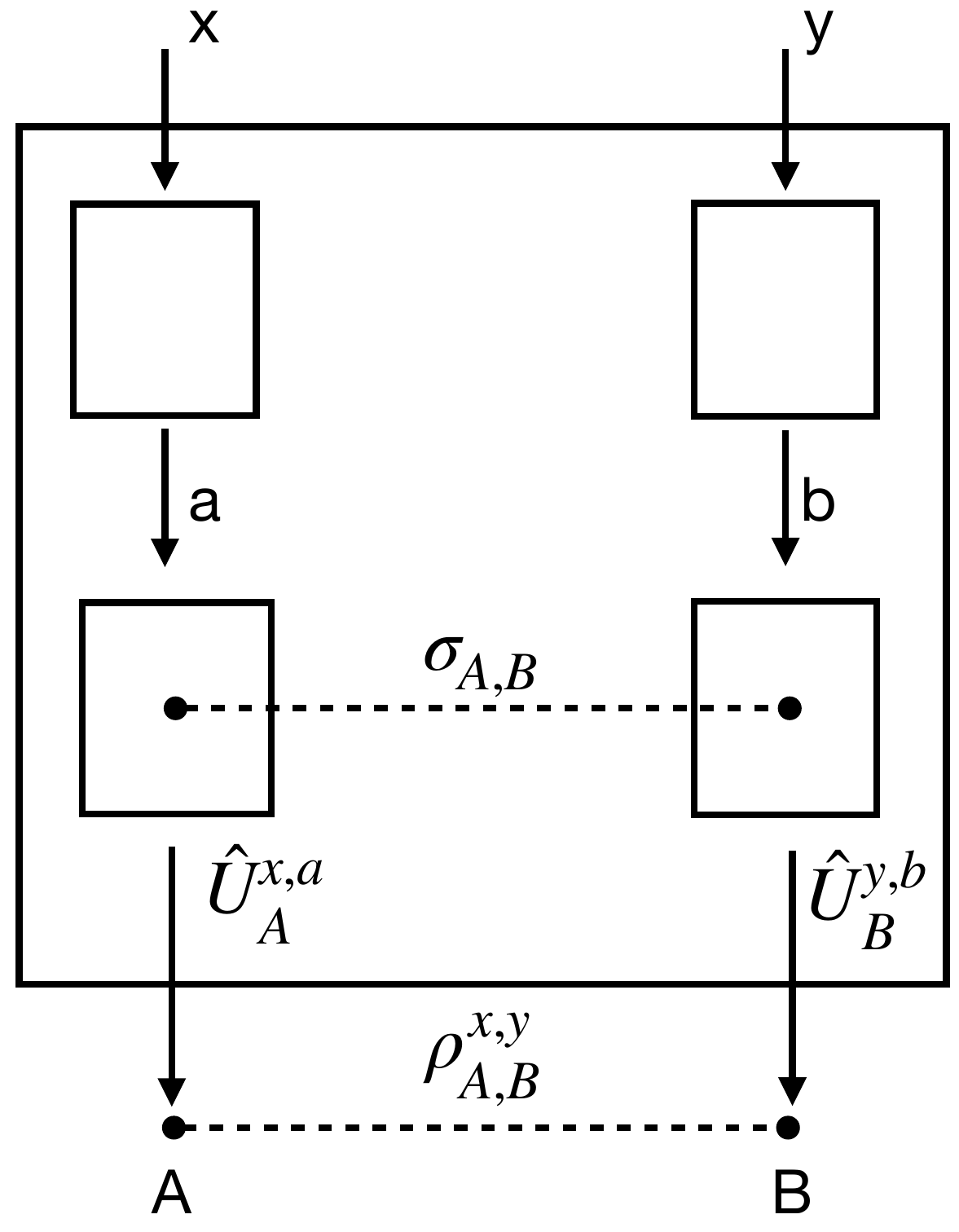}
\caption{Example of a non-genuine C-Q box, constructed from a C-C box (which takes inputs x,y and outputs a,b), a pre-shared entangled state $\sigma_{A,B}$, and some local unitaries $\hU^{x,a}_A$ and $\hU^{y,b}_B$ that are applied to $\sigma_{A,B}$ to generate the desired state $\rho_{A,B}^{x,y}$. In general ancillas may be also included inside the box.}
\label{Fig:CQBoxes2}
\end{figure}

Clearly, some C-Q boxes could be built by simply having inside pre-shared entangled states and performing local unitary evolutions depending on the local inputs. Then there are more sophisticated ways in which some C-Q boxes could be constructed by simpler, already known objects. For example some may be constructed by having inside 
classical input - classical output (C-C) non-signalling
boxes (such as, for example PR boxes), feeding the inputs into them and using their outputs to determine local unitaries acting on  pre-shared entangled states (see Fig \ref{Fig:CQBoxes2}). But are there any C-Q boxes that cannot be decomposed in this way? This is the specific question we address here. 

Regardless of the answer, there are several reasons for considering such boxes. Quantum states have properties that are not captured by the formalism of the standard classical input - classical output (C-C) non-signalling boxes. While C-C non-signalling boxes can present correlations stronger than those allowed by quantum mechanics, their dynamics are far more limited than that of the quantum states \cite{short,barrett2010}. In particular, while nonlocality swapping (algebraically described by entanglement swapping) is possible in quantum mechanics, it is not possible for C-C boxes \cite{short}.  Obviously then, if genuine C-Q non-signalling boxes exist, they will extend the range of non-local phenomena that we thought to be possible in a non-deterministic world, while being consistent with relativity. On the other hand, if genuine C-Q boxes do not exist, another set of interesting questions follow. First, what are the resources needed to implement them via C-C boxes and pre-shared entanglement? As we will show, in some cases it seems that these resources need to be extremely large. Second, and more important, why do genuine C-Q boxes not exist?  Why is it that the most nonlocal non-signalling device with quantum output is an ordinary pairing of a C-C non-signalling device and pre-shared entanglement?

As for the answer, the question of the existence of genuine C-Q boxes is still open. 

The main result of our paper is that {\it all bi-partite C-Q boxes whose outputs are pure states are not genuine}, ruling out one of the major class of situations. To put the result in context, most of the major results concerning non-locality in the traditional classical input-classical output (C-C) scenario, starting with the very discovery by J. Bell of the existence of nonlocality, have been obtained in the bi-partite, pure state situations. One would have expected that if the extension of non-locality to classical input-quantum output situations (C-Q) brings something new, this should already be evident from the extension of this simpler class which, as we show, is not the case.

At the same time, it is also the case that in the C-C scenario mixed states and multi-partite situations manifest qualitatively new phenomena with respect to the bi-partite, pure state situation. It is thus conceivable that the C-Q scenario in these more general situations does contain genuine C-Q effects that cannot be simulated by using C-C correlations. Here we only have partial results: (i) For bi-partite C-Q boxes with mixed state outputs we shall present various paradigmatic examples of boxes which turn out to be non-genuine. (ii) For multi-partite situations the no-signalling constraints  appear so strong that we conjecture all pure state C-Q boxes are non-genuine, and additionally that most of them are LOSE, i.e. can be implemented using only entanglement and local operations, without any C-C boxes which go beyond quantum mechanics.

Along the way of proving the above results, our paper sets up the general strategy for investigating C-Q non-locality and exposes basic structures of C-Q boxes. As a by-product we also draw the attention to a novel class of C-C boxes, extensions of the PR boxes which have been a main tool in the study of non-locality.  

Finally we show that even some very simple non-genuine C-Q boxes require large amounts of C-C nonlocal correlations in order to simulate them. This shows the power of C-Q boxes, and sheds light on how quantum mechanics and correlations beyond quantum mechanics fit together.

Zooming out,  we note that while in this paper we focus specifically on the question of whether or not C-Q boxes can be simulated by C-C boxes plus shared entanglement, one could consider other known resources, (for example some GPTs), and ask whether they can be used to simulate all C-Q boxes. This further question would really come into play if we find C-Q boxes that cannot already be simulated by C-C boxes + entanglement.

\section{Non-genuine C-Q boxes}

Obviously, some C-Q boxes could be obtained by a combination of pre-shared entangled quantum states and a classical-classical non-signalling box, which are encapsulated in a bigger box so that from the outside one doesn't see this combination, as illustrated in Fig \ref{Fig:CQBoxes2}.

Specifically, the inputs $x$ and $y$ are plugged in the C-C box, which gives outputs $a$ and $b$.  Then Alice applies some unitary transformation  $\hU^{x,a}_A$ to her quantum particle depending on her input $x$ and the C-C box output $a$ and finally outputs her quantum particle out of the bigger box. Bob follows a similar procedure. Internal ancillas could also be added. From the outside this combination looks as a C-Q box. We call these ``non-genuine" classical-quantum boxes.

Note that the presence of a C-C box in addition to the shared entanglement is non-trivial and adds something, only when the correlations generated by it go beyond quantum mechanics; otherwise everything could be absorbed into the shared entangled state.

\section{Examples of boxes with pure states outputs}

We present a few examples of boxes, of increasing complexity, which turn out to be non-genuine.  This will show the main ideas of how to create the desired boxes and show that they are non-genuine.

\subsection{QM correlations are not enough}
\label{simpleNonGenuineCQBox}
As a first example consider inputs $x,y=0,1$ and the C-Q box defined by  \beq \begin{split} x \cdot y = 0 \rightarrow \frac{1}{\sqrt{2}} ( \ket{0} \ket{0} + \ket{1} \ket{1} ) \\
x \cdot y = 1 \rightarrow \frac{1}{\sqrt{2}} ( \ket{0} \ket{1} + \ket{1} \ket{0} )
\end{split} \eeq
where $\ket{0}\ket{1}$ is short for $\ket{0}_A\ket{1}_B$. 

Note that this box {\it cannot} be created with only shared entanglement and no other non-local resources such as C-C boxes, as one can make the PR box 
\beq (a - b) \bmod{2} = x\cdot y \eeq
from this by measuring in the $0/1$ basis, and the PR box (which is one example of a C-C box) is known to give stronger than quantum correlations.

This box however is a non-genuine C-Q one since it can be simulated by the PR box above plus the  pre-shared maximally entangled state $\ket{\Phi} = \frac{1}{\sqrt{2}} ( \ket{0} \ket{0} + \ket{1} \ket{1} ) $.  Alice simply needs to apply the bit flip operator $U_A$ (which flips $\ket{0}$ to $\ket{1}$ and vice versa) to $A$ whenever the PR box gives $a=1$.  Similarly Bob applies the same bit flip operator $U_B$ to $B$ whenever $b=1$.  When $x\cdot y=0$ the PR box gives $a=b$, so Alice and Bob will either both flip or neither flip their qubits, and both operations leave $\ket{\Phi}$ unchanged.  When $x\cdot y=1$ only one of Alice or Bob will flip their bits and we will have $\frac{1}{\sqrt{2}} ( \ket{0} \ket{1} + \ket{1} \ket{0} )$ as desired.

\subsection{Sign flip}
Next consider inputs $x,y=0,1$ and the box which outputs states $\ket{\Psi^{x,y}}$ according to
\beq \ket{\Psi^{x,y}}=\alpha\ket{0}\ket{0}+ \beta e^{i\pi x\cdot y} \ket{1}\ket{1}. \eeq
This can be simulated by a PR box $(a-b)\bmod{2}=x\cdot y$. Alice and Bob apply $\hU^a_A$ and $\hV^b_B$ respectively to the initial state $\ket{\Psi}=\alpha\ket{0}\ket{0}+\beta\ket{1}\ket{1})$ where
\beq \begin{aligned} 
\hU^a_A\ket{0}_A &= \ket{0}_A & \hV^b_B\ket{0}_B &= \ket{0}_B \\ 
\hU^a_A\ket{1}_A &= e^{i \pi a} \ket{1}_A & \hV^b_B\ket{1}_B &= e^{-i \pi b} \ket{1}_B
\end{aligned} \eeq 
which leads to
\beq \begin{split} \ket{\Psi}\rightarrow  \hU^a_A\hV^b_B\ket{\Psi} & =\alpha\ket{0}\ket{0}+ \beta e^{i\pi (a-b)}\ket{1}\ket{1}\\
& =\alpha\ket{0}\ket{0}+ \beta e^{i\pi ((a-b) \bmod{2})}\ket{1}\ket{1}\\
& =\alpha\ket{0}\ket{0}+\beta e^{i\pi x\cdot y}\ket{1}\ket{1} \end{split} \eeq
which is the desired state.

\subsection{Phase change} 
\label{PhaseChange}

The sign change in the previous example can be generalized to an arbitrary rational phase parameterized by $\theta$:
\beq 
\label{phaseChangeBox}
\ket{\Psi^{x,y}}=\alpha\ket{0}\ket{0}+\beta e^{i 2\pi\theta x\cdot y} \ket{1}\ket{1}. 
\eeq

In the previous case $\theta=1/2$. Suppose now $\theta = 1/4$.  It is simple to see that using a standard PR box one cannot implement this C-Q box. Does this mean that this box is genuine C-Q?  No. To decide that a C-Q box is genuine we need to show that there is {\it no} way to construct it by using a standard classical-classical box and using its outputs to implement appropriate local unitary operations on a pre-shared entangled quantum state. In our case it turns out that this is possible.

The desired C-Q box can be constructed by the use of a C-C box which takes inputs $x,y=0,1$ and gives outputs $a,b=0,1,2,3$ according to 
$(a-b) \bmod{4} = x \cdot y$,
with all pairs of outcomes that respect this constraints being given with equal probability (1/4 in this case).  Alice then performs the rotation $\hU^a_A\ket{1}_A=e^{i2\pi a/4}\ket{1}_A$ and Bob $\hV^b_B\ket{1}_B=e^{i2\pi (-b)/4}\ket{1}_B$. This gives 
\beq \begin{split} \ket{\Psi} & \rightarrow \alpha\ket{0}\ket{0}+ \beta e^{i2\pi (a-b)/4}\ket{1}\ket{1}\\
& =\alpha\ket{0}\ket{0}+\beta e^{i2\pi x\cdot y / 4}\ket{1}\ket{1} \end{split} \eeq
as desired.  

For other rational values of $\theta$, i.e. $\theta = \frac{m}{n}$ where $m$ and $n$ are integers, we can use a similar box with $n$ dimensional outputs giving $(a-b) \bmod{n} = x \cdot y$ and rotations $\hU^a_A\ket{1}_A = e^{i2\pi a m/n}\ket{1}_A$ and $\hU^b_B\ket{1}_B = e^{i2\pi (-b) m/n}\ket{1}_B$. 

\subsection{Phase change boxes and use of resources} 
\label{phaseChangeResources}

As we have seen before, any phase change box with $\theta$ equal to a rational number, $\theta = \frac{m}{n}$, can be realised by a C-C box and unitary transformations, provided that we use a C-C box with $n$ outcomes. When $n$ is large, this C-C box is a large amount of non-local resources.  One might ask whether there is a more efficient way to implement this C-Q box.  However we shall show here that this is the most efficient possible implementation, both in terms of C-C box and entanglement.

Proving that to implement a phase change box with $\theta = \frac{1}{n}$ we require a C-C box with $a$ and $b$ each having $n$ outcomes, defined by $(a-b) \bmod{n} = x \cdot y$, proceeds as follows.  

First, assume that the procedure is implemented by starting with $\ket{\Psi} = \alpha \ket{0}\ket{0}+ \beta \ket{1}\ket{1}$, and applying unitary operations $\hU^a_A$ and $\hV^b_B$ when the C-C box outputs $a$ and $b$. For simplicity we consider the case when $\alpha\neq \beta$, so the only operations that are allowed are phase shifts. Furthermore, we can take the $\hU^a_A$ to be different for different $a$, since otherwise we can use a simplified C-C box by merging them together, and similarly for Bob.

For the case $(x,y)=(0,0)$, when $a=0$, which occurs with some probability $p_0$, Alice applies $\hU^0_A$.  To generate $\ket{\Psi}$ Bob must apply $(\hU^0_B)^*$, i.e. the complex conjugate.  We can label his C-C box outcome $b=0$ in this case.

The next step is to realise that for the inputs $(x,y)=(0,0)$ the pair of outputs $(a,b)=(0,1)$ is redundant as Bob will need to apply the same unitary $(\hU^0_B)^*$ as for $(a,b)=(0,0)$. Hence, for the inputs $(x,y)=0,0$ we will only consider $a=0$ to be paired with $b=0$. Similarly for $a=j$ when Alice applies $\hU^j_A$, Bob must apply $(\hU^j_B)^*$ and we can label his outcome $b=j$.  So we can describe the C-C box in this $(x,y)=(0,0)$ case as outputting $a=b=j$ with probability $p_j$.

Due to no-signalling of the C-C box from Bob to Alice, comparing $(x,y)=(0,1)$ to $(0,0)$ we see that in the case $(0,1)$ Alice must receive outcome $a=j$ with the same probability $p_j$ as for the $(0,0)$ inputs, upon which she will apply $\hU^j_A$.  To keep $\ket{\Psi}$ unchanged the C-C box must output $b=a$, so that Bob applies $(\hU^j_B)^*$.  Similarly due to no-signalling from Alice to Bob, comparing $(x,y)=(1,0)$ to $(0,0)$ we see that in the case $(1,0)$ the C-C box must output $b=j$ with probability $p_j$, and to generate $\ket{\Psi}$ give $b=a$.  

The important difference comes when considering the inputs $(x,y)=(1,1)$. Comparing $(1,0)$ to $(1,1)$ it is still the case that for  $(x,y)=(1,1)$ Alice must receive outcome $a=j$ with probability $p_j$ and apply $\hU^j_A$.  And comparing $(0,1)$ to $(1,1)$ we see that for $(1,1)$ Bob must receive outcome $j$ with probability $p_j$ and apply $(\hU^j_B)^*$.

However to generate the state 
\beq
\ket{\Psi^{1,1}} = \alpha \ket{0}\ket{0} + \beta e^{i 2 \pi / n} \ket{1}\ket{1}
\eeq
we need the C-C box to pair up $a$ and $b$ so that 
\beq
(\hU^b_B)^* \hU^a_A \ket{\Phi} = \ket{\Psi^{1,1}} = (\hSigma^{2/n}_z)_A \ket{\Phi},
\eeq
where $\hSigma_z$ is the Pauli matrix which does 
\beq \begin{split}
\ket{0} &\rightarrow \ket{0} \\
\ket{1} &\rightarrow e^{i \pi} \ket{1}.
\end{split} \eeq

Since the unitaries $\hU^a_A$ and $\hU^b_B$ only apply phases, i.e. are powers of $\hSigma_z$ themselves, this implies 
\beq 
\hU^a_A = (\hSigma^{2/n}_z)_A \hU^b_A.
\eeq
i.e. for any $j$, if $\hU^j_A$ is one of the unitaries used in implementing the C-Q box, then $(\hSigma^{2/n}_z)_A \hU^j_A$ is another one.  The smallest set of $\hU^j_A$ which has this property is the one where $j=0..n-1$ and 
\beq
    \hU^j_A = (\hSigma^{2j/n}_z)_A.
\eeq
Given this set, we can create the desired state by pairing together $a$ and $b$ as 
\beq (a-b) \bmod{n} = x \cdot y\eeq
which is a generalisation of the standard PR box that can be written also as $(a-b) \bmod{2} =x\cdot y$.

This proves that if we implement the C-Q box in Eq. \eqref{phaseChangeBox} with $\theta = \frac{1}{n}$, using an initial $\ket{\Psi}$, local unitaries and a C-C box, then the C-C box needs to have dimension $n$ and be of the form $(a-b) \bmod{n} = x \cdot y$.  

Finally one could imagine starting with a different initial shared entangled state, and hope that that will allow us to reduce the C-C box resources.  We cover this case in Appendix \ref{appendixResourcesProof}.
Thus any implementation of this C-Q box requires a C-C box with $n$ outputs. 

\subsection{C-Q Box with maximally entangled outputs for each input.} 

Consider a C-Q box with maximally entangled outputs for each input. 

Any arbitrary maximally entangled state can be written by acting on, say, Alice's side with a unitary on a standard maximally entangled state. Using this representation we can define an arbitrary C-Q box with maximally entangled outputs as:
\beq 
\label{boxWithUnitaryOnAll}
\begin{split} &\ket{\Psi^{0,0}}=\ha_A\ket{\Psi^-}\\  &\ket{\Psi^{0,1}}=\hb_A\ket{\Psi^-}\\ &\ket{\Psi^{1,0}}=\hc_A\ket{\Psi^-}\\ &\ket{\Psi^{1,1}}=\hd_A\ket{\Psi^-},  \end{split} \eeq
where $\ket{\Psi^-} = \frac{1}{\sqrt{2}}(\ket{0}\ket{1}-\ket{1}\ket{0})$.

To show that this box is non-genuine C-Q we will show how to implement it using two steps, of which only the first uses a C-C box.  First we will show how to implement a simplified version of the above box where $\ha_A = \hb_A = \hc_A = \mathbf{1}$ and $\hd_A = \hat{U}_A$ for an arbitrary $\hat{U}$.

This can be done by noting that any unitary on a $2$ dimensional system can be viewed in the Block sphere as a rotation by $\theta$ around a given axis, and the phase change implemented above is exactly such a rotation around the 0/1 axis.  As $\ket{\Psi^-}$ is rotationally symmetric we can always write it as $\frac{1}{\sqrt{2}} \left( \ket{\vec{n}} \ket{\vec{-n}} - \ket{\vec{-n}} \ket{\vec{n}} \right)$ where $\vec{n}$ is the axis of rotation, and then use the the phase change box from the previous section to apply the rotation around that axis.  

To implement the more general box in Eq. \eqref{boxWithUnitaryOnAll} we shall next apply some local unitary operations.
Alice applies $\ha_A$ when $x=0$ and $\hc_A$ when $x=1$, and Bob does nothing when $y=0$ and $\hb^{\dagger}_B \ha_B$ when $y=1$.  This gives us 
\beq 
\begin{split} 
&\ket{\Psi^{0,0}}=\ha_A\ket{\Psi^-}\\  &\ket{\Psi^{0,1}}= \ha_A \hb^{\dagger}_B \ha_B \ket{\Psi^-}\\ 
&\ket{\Psi^{1,0}}=\hc_A\ket{\Psi^-}\\ &\ket{\Psi^{1,1}}=\hc_A \hb^{\dagger}_B \ha_B \hU_A \ket{\Psi^-}.  
\end{split} \eeq

Now note that due to the rotational symmetry of $\ket{\Psi^-}$ if Alice and Bob act on their particles with the same unitary operator $\hU$, the state remains unchanged. So, in particular 
\beq \begin{split} \hU_A\hU_B\ket{\Psi^-} &=\ket{\Psi^-}\\
\hU_B\ket{\Psi^-} &=\hU_A^{\dagger}\ket{\Psi^-}. \end{split} \eeq
Thus 
\beq 
\begin{split} \ket{\Psi^{0,1}} &= \ha_A (\hb^{\dagger}_A \ha_A)^{\dagger} \ket{\Psi^-}\\  
&= \ha_A \ha^{\dagger}_A \hb_A \ket{\Psi^-} \\
&= \hb_A \ket{\Psi^-}, 
\end{split} \eeq
and
\beq 
\begin{split} \ket{\Psi^{1,1}} &= \hc_A \hU_A \hb^{\dagger}_B \ha_B \ket{\Psi^-}\\ 
&= \hc_A \hU_A (\hb^{\dagger}_A \ha_A)^{\dagger} \ket{\Psi^-} \\
&= \hc_A \hU_A \ha^{\dagger}_A \hb_A \ket{\Psi^-}. 
\end{split} \eeq

We can thus achieve $\ket{\Psi^{1,1}} = \hd_A\ket{\Psi^-}$ by setting $\hU_A = \hc^{\dagger}_A \hd_A \hb^{\dagger}_A \ha_A$. 

Thus we have shown how to implement our C-Q box with maximally entangled outputs in terms of C-C boxes and local unitary operations.

Finally, we note that the protocol involves creating a phase change box to implement $U$. The resources used are therefore the ones necessary to implement the phase change, which depend on what the phase is as described in section \ref{phaseChangeResources}. They could be very large, and if the phase is irrational we will not be able to implement $U$ exactly using this method.  Instead we would need one of the methods we shall discuss later in sections \ref{MainIdea} and \ref{irrationalPhaseChange}.

\subsection{Higher Dimensions} 
\label{higherDimensionExample}
We will now consider C-Q boxes with more inputs, say $x=0,1$ and $y=0,1,2$. This gives the hope to find a genuine C-Q box. The idea is that there have been already many constraints due to non-signalling in implementing via C-C + entanglement a C-Q box on the subset of inputs $x,y=0,1$ and that the input $y=2$ when paired with $x=0,1$ will add supplementary constraints that can no longer be fulfilled. 

Consider the box \beq \begin{split} \ket{\Psi^{0,0}} = \ket{\Psi^{0,1}} = \ket{\Psi^{0,2}} = \ket{\Psi^{1,0}} &= \frac{1}{\sqrt{2}} ( \ket{0} \ket{0} + \ket{1} \ket{1} ) \\
\ket{\Psi^{1,1}} &= \hSigma_z^{1/2} \ket{\Psi^{0,0}} \\
\ket{\Psi^{1,2}} &= \hSigma_x \ket{\Psi^{0,0}},
\end{split} \eeq
where $\hSigma_z$ is the usual Pauli operator
\beq \begin{split}
    \hSigma_z \ket{0} &= \ket{0} \\
    \hSigma_z \ket{1} &= - \ket{1},
\end{split} \eeq
and $\hSigma_x$ flips the bits
\beq \begin{split}
    \hSigma_x \ket{0} &= \ket{1} \\
    \hSigma_x \ket{1} &= \ket{0}.
\end{split} \eeq

This is the same as phase changes we handled in section \ref{PhaseChange} for $(x,y) \in \{(0,0), (0,1), (1,0), (1,1)\}$.  Thus we could construct that part from a pre-shared $\ket{\Psi^{0,0}}$ using the C-C box $(a-b) \bmod{4} = x.y$,  Alice applying the unitary operator 
\beq 
\label{sigmaZUnitaries}
\{ \mathbf{1},  \hSigma_z^{1/2}, \hSigma_z^{2/2}, \hSigma_z^{3/2}\} 
\eeq 
to $A$ when $a = 0,1,2,3$ respectively, and Bob applying the inverse operator 
\beq 
\{ \mathbf{1},  \hSigma_z^{-1/2}, \hSigma_z^{-2/2}, \hSigma_z^{-3/2}\} 
\eeq
to $B$ when he sees $b = 0,1,2,3$ respectively.

However in order to create $\ket{\Psi^{1,2}}$ when $(x,y) = (1,2)$ we need something which allows us to flip the bit, e.g. $\hSigma_x$.  The no-signalling condition means that any C-C box we use must have the same set of outputs $a$ for the $(x,y)$ cases $(1,1)$ and $(1,2)$.  So it seems we have to add $\hSigma_x$ to the unitaries in Eq. \eqref{sigmaZUnitaries}.  However since $\hSigma_x$ doesn't commute with $\hSigma_z$, it looks likely to break the state we carefully constructed for $(x,y)=(1,1)$.  The solution is to use a new C-C box, described below, which has 8 outputs for $a$ and $b$ instead of 4.  To make the desired states Alice will perform $\hU_a$ when the C-C box outputs a, and Bob will perform $\hU_b^*$ (the complex conjugate) when the C-C box outputs b, where $\hU_i$ is defined as
\begin{align}
\hU_0 &= \mathbf{1}, &\hU_1 &= \hSigma_z^{1/2}, &\hU_2 &= \hSigma_z^{2/2}, &\hU_3 &= \hSigma_z^{3/2}, \nonumber\\
\hU_4 &= \hSigma_x, &\hU_5 &= \hSigma_z^{1/2} \hSigma_x, &\hU_6 &= \hSigma_z^{2/2} \hSigma_x, &\hU_7 &= \hSigma_z^{3/2} \hSigma_x.
\end{align}

When $(x,y) \in \{ (0,0), (0,1), (0,2), (1,0) \}$ the C-C box outputs $a=b$, which gives 
\beq \hU^A_a (\hU^B_a)^* \ket{\Psi^{0,0}} = \ket{\Psi^{0,0}} \eeq 
(see appendix \ref{appendixStateInvariance} for a more detailed proof).  

To make $\ket{\Psi^{1,1}}$, the C-C box outputs the pairs 
\beq
\{(1,0), (2,1), (3,2), (0,3), (5,4), (6,5), (7,6), (4,7)\}.
\eeq
In other words we pair together the cases where both a and b are less than 4, and then the cases where both a and b are at least 4.  This works as the state is invariant under $\hat{\sigma_x}$ performed by Alice and Bob simultaneously.

To make $\ket{\Psi^{1,2}}$, we instead pair together $a$ and $b$ as 
\beq
\{(4,0), (7,1), (6,2), (5,3), (0,4), (1,7), (2,6), (3,5)\}.
\eeq  
It's straightforward to check this works as desired.  The box is non-signalling since the outputs $a$ and $b$ occur with the same probability independent of $x$ and $y$.

\section{General Pure State Theorem}
\label{generalPureStateTheorem}

Here we shall show that any C-Q non-signalling boxes which output a set of bi-partite pure states are non-genuine.  We build the proof by first showing how to deal with the case of outputs that are maximally entangled states, then the case of non-maximally entangled states - the two cases being different in the constraints of the unitaries used when trying to implement them via C-C boxes - and finally to the general C-Q boxes.

\subsection{The main idea}
\label{MainIdea}

First we show the main idea applied to a simple case.

Suppose we want our box to output:
\beq 
\label{generalQubitCcBox}
\ket{\Psi^{x,y}}=(\hat{\alpha}_A)^{x\cdot y} \frac{1}{\sqrt{2}}(\ket{0}\ket{0}+\ket{1}\ket{1}), 
\eeq where $\ha_A$ is an arbitrary unitary on A; taking it to the power $x\cdot y$ with $x,y=0,1$ means that we only apply it when $x=y=1$, and in all other cases it is the identity.  We could achieve this using the method described in Section \ref{PhaseChange}, however here we present a more powerful method which allows us to handle many more cases.  Again we shall start with an entangled state $\ket{\Phi^+}=\frac{1}{\sqrt{2}} ( \ket{0} \ket{0} + \ket{1}\ket{1} )$.  

Thus far we have considered C-C boxes with integer outputs, from $0$ to $n-1$, and used them to determine one of a finite set of unitaries.  However the group of all possible unitary operations Alice could apply to her qubit is SU(2), which is an infinite set.  One way to parameterize it is to use the local isomorphism of SU(2) to SO(3), the group of rotations in 3-dimensional space, and to define the axis of rotation by the Euler angles $(\phi, \theta)$, and the size of the rotation by $\psi$, where $0 \le \psi < 4 \pi$.  The size of the rotation goes up to $4 \pi$, a double rotation, as the map from SU(2) to SO(3) is a double covering: it takes two complete rotations to return to the initial state.  We thus use a C-C box which takes inputs $x,y=0,1$ and outputs $a$ and $b$, where $a$ is a triple $(\phi_a, \theta_a, \psi_a)$ in order to label an element of SU(2), and similar for $b$.  These outputted angles could be analog dials which rotate in order to point to the exact angle\footnote{In practice we could use a digital output which gives a decimal expansion of the angle to a fixed number of digits of precision.  This introduces a rounding error compared to the exact angle, which can be made as small as desired by adding more digits of precision. The rounding can be implemented at a local level, which ensures there is no possibility of introducing signalling into the C-C box.}.

Our C-C box is arranged so that for any fixed $(x,y)$, it outputs $b$ such that $\hU^b$ is distributed according to the Haar measure on SU(2): essentially $\hU^b$ is a unitary chosen uniformly at random. When the box outputs $b$ on Bob's side, it outputs $a$ on Alice's side so that $\hU^a = (\hat{\alpha}^*)^{x\cdot y} \hU^b$.  For each fixed $(\hat{\alpha}^*)^{x\cdot y}$, this ensures that $\hU^a$ is also distributed according to the Haar measure (this follows straightforwardly from the left-translation invariance in the definition of the Haar measure).  This is a generalization of the intuition that if one starts with a distribution which chooses a rotation uniformly at random, and then rotates the whole distribution through a fixed angle, the new distribution will be the same as the old one.

Since the distributions of $a$ and $b$ are random, which in particular means that they are independent from the inputs $x$ and $y$, our C-C box is no-signalling.   

We now build a C-Q box by starting with $\ket{\Psi^{x,y}}$ and applying on Alice's side $(\hU^a_A)^*$ and on Bob's side $\hU^b_B$.  Since all unitaries on Alice and Bob's sides are possible, the final C-Q box depends upon how $a$ and $b$ are correlated.  For $x \cdot y=0$, the C-C box correlates $a$ and $b$ so that Alice applies the complex conjugate of Bob's unitary.  i.e. when Bob does $\hU^b_B$, Alice does $(\hU^a_A)^* = (\hU^b_A)^*$.  This leaves $\ket{\Phi^+}$ unchanged as $(\hU^b_A)^* \hU^b_B \ket{\Phi^+} = \ket{\Phi^+}$ (see Appendix \ref{appendixStateInvariance}).  For $x \cdot y=1$, the C-C box correlates $a$ and $b$ so that when Bob does $\hU^b_B$, Alice does $(\hU^a_A)^* = \hat{\alpha}_A (\hU^b_A)^*$.  This gives $\ket{\Phi^+} \rightarrow \hat{\alpha}_A \ket{\Phi^+}$, and so we have implemented the box in Eq. \eqref{generalQubitCcBox}. 

In terms of resources this is very expensive: we have 3 real parameters describing the C-C box we use to specify $a$ and $b$, but we expect that for any particular $\hat{\alpha}_A$ there will be a simpler C-C box which allows us to implement this C-Q box.

\subsection{Maximally entangled pure states}
\label{Maximally Entangled Pure States}

Now we generalize the main idea to output states of arbitrary dimension $n$, and the inputs $x$ and $y$ to arbitrary dimension.  Recall that in section \ref{higherDimensionExample} we showed an example of a particular C-Q box with higher input dimensions, which had additional constraints which made it more difficult to implement.  Nevertheless we found a method to implement it using a C-C box and shared entanglement.  Here we shall use the Haar measure idea from the previous section to generalize this to all bi-partite C-Q boxes outputting pure states.  We show that all such boxes are non-genuine.

Consider the box
\beq \begin{split} \ket{\Psi^{x,y}} &= \hU^{x,y}_A \ket{\Phi^{n+}}, \\
\text{where } \ket{\Phi^{n+}} &= \frac{1}{\sqrt{n}} \sum_{i=0}^{n-1} \ket{i}_A \ket{i}_B, \end{split} \eeq
and $n$, $x$ and $y$ are non-negative integers.  Note that all maximally entangled pure states of dimension $n$ can be obtained from any one of them by local rotations $\hU^{x,y}_A$ on A, so this covers a large class of non-signalling pure states.  

To implement this box we follow the same idea: start with the pre-shared state $\ket{\Phi^{n+}}$, and use a C-C box which distributes $a$ and $b$ according to the Haar measure over SU(n), and which correlates them so that when Bob does $\hU^b_B$, Alice does $\hU^a_A = \hU^{x,y}_A (\hU^b_A)^*$.  This works as desired. 

\subsection{Non-maximally entangled pure states}
\label{NonMaximalPureStates}

Next we show that all bi-partite C-Q boxes which output non-maximally entangled pure states, where the weights in the Schmidt decomposition are all different from one another, are non-genuine.  

First we show that any such box can be written in the form
\beq 
\label{nonMaximalPureStateForm}
\begin{split}
\ket{\Psi^{x,y}} &= \hU^x_A \hat{V}^y_B \hat{W}^{x,y}_A \ket{\Phi^n}, \\
\text{where } \ket{\Phi^n} &= \sum_{i=0}^{n-1} \sqrt{p_i} \ket{i}_A \ket{i}_B, \\
\hat{W}^{x,y}_A \ket{i}_A &= e^{i 2\pi \alpha^{x,y}_i} \ket{i}_A, \\
\sum_{i=0}^{n-1} p_i &= 1,  
\end{split} \eeq
$\hU^x_A$ and $\hat{V}^y_B$ are local unitaries, and the $p_i$ are distinct.
In other words we can relate the various output states of the box by $\hat{W}^{x,y}_A$ which applies phases parameterized by $\alpha^{x,y}_i$ in the computational basis $\ket{i}$, and local unitaries $\hU^x_A$ and $\hat{V}^y_B$.

Why is this?  Starting with $(x,y)=(0,0)$, by choosing our basis appropriately we can write the output state in its Schmidt decomposition 
\beq
\ket{\Psi^{0,0}} = \ket{\Phi^n}.
\eeq
Next, for $(x,y)=(1,0)$, no-signalling from Alice to Bob implies that Bob's reduced density matrix is the same for $(0,0)$ and $(1,0)$, i.e.
\beq
Tr_A \left( \ket{\Psi^{1,0}} \bra{\Psi^{1,0}} \right) = Tr_A \left( \ket{\Psi^{0,0}} \bra{\Psi^{0,0}} \right).
\eeq
Since the state is bi-partite and pure, this implies that we can relate the two output states via a local unitary on Alice's side: call this $\hU_A^1$.  i.e.
\beq
\ket{\Psi^{1,0}} = \hU_A^1 \ket{\Psi^{0,0}}.
\eeq
Similarly, no-signalling from Bob to Alice implies that there exists a unitary $\hV_B^1$ such that
\beq
\ket{\Psi^{0,1}} = \hV_B^1 \ket{\Psi^{0,0}}.
\eeq
Next consider the case $(1,1)$.  We can either compare this with $(0,1)$ giving
\beq
\ket{\Psi^{1,1}} = \hS_A^1 \ket{\Psi^{0,1}},
\eeq
or compare it with $(1,0)$ giving
\beq
\ket{\Psi^{1,1}} = \hT_B^1 \ket{\Psi^{1,0}}.
\eeq
These two ways of writing the same state imply
\beq
\hS_A^1 \hV_B^1 \ket{\Phi^n} = \hT_B^1 \hU_A^1 \ket{\Phi^n}.
\eeq
This implies that 
\beq
\hS_A^1 = \hU_A^1 \hW_A^{1,1},
\eeq
for some $\hW_A^{1,1}$ which applies a phase in the Schmidt basis.  Therefore 
\beq
\ket{\Psi^{1,1}} = \hU_A^1 \hV_B^1 \hW_A^{1,1} \ket{\Phi^n}.
\eeq
Similar logic for the other values of $(x,y)$ generalizes this to Eq. \eqref{nonMaximalPureStateForm}.

Now that we have determined the general form of all C-Q boxes with distinct $p_i$, we will show that they are non-genuine.  We can implement $\hat{W}^{x,y}_A$ easily using the fact that the unitaries applying the different phases commute by following the final protocol in section \ref{PhaseChange}.  After that we just apply the local unitaries.  This is in fact easier to implement than the maximally entangled case, as the no-signaling constraint forces $\hat{W}^{x,y}_A$ to only apply phases, whereas in the maximally entangled case it can be an arbitrary unitary.

\subsection{General pure states}

Finally, to cover all possible sets of pure states, we consider what happens when some of the $p_i$ are equal.  In that case we can view any subset of $\ket{i}$ where $p_i$ are equal as forming a maximally entangled subspace, noting that there may be several of these sub-spaces.  Then we can create the desired set of states for each of those sub-spaces using the method in section \ref{Maximally Entangled Pure States}, then implement the necessary operations for those sub-spaces vs the others using the methods in section \ref{NonMaximalPureStates}.

Thus we have shown how to implement any non-signalling set of pure bipartite states.

\section{Mixed States}

We now look at C-Q boxes which produce quantum states which may be mixed or pure.  One might think that this is a simple extension of the pure state case, as every mixed state can be written as a probabilistic mixture of pure states.  However that is not the case, as we do not know of any guarantee that a set of non-signalling mixed states can be written as a probabilistic mixture of non-signalling sets of pure states.

Thus mixed state C-Q boxes have a lot more freedom, and in general some of them may be genuine.  Below, however, we show that a reasonably large class of such boxes are in fact non-genuine.  The general case remains open.

\subsection{Maximally Disordered Qubits}

We start by showing how to create any C-Q box, $\rho_{AB}^{x,y}$, outputting states of 2 qubits whose local density matrices are maximally disordered, i.e. 
\beq \begin{split}
Tr_A(\rho_{AB}^{x,y}) &= \mathbf{1}_B / 2 \\
Tr_B(\rho_{AB}^{x,y}) &= \mathbf{1}_A / 2.
\end{split} \eeq

Examples of such states are the maximally entangled Bell states, product states of 2 completely uncertain qubits, and a classical mixture of $\ket{0}\ket{0}$ and $\ket{1}\ket{1}$.  

It is shown in \cite{DensityMatrix} that any 2 qubit density matrix $\rho$ with both single qubit reduced density matrices maximally disordered can be written as a probabilistic mixture of Bell states with local unitary operators applied to $A$ and $B$, i.e.
\beq
\rho = \hU_A \hV_B \left( \sum_{i=0}^3 p_i \ket{\Phi_i}\bra{\Phi_i} \right) \hU_A^{\dagger} \hV_B^{\dagger},
\eeq
where
\beq \label{bellStates} \begin{split}
\ket{\Phi_0} &= \frac{1}{\sqrt{2}}(\ket{0}\ket{0} + \ket{1}\ket{1}),\\
\ket{\Phi_1} &= \frac{1}{\sqrt{2}}(\ket{0}\ket{1} + \ket{1}\ket{0}),\\
\ket{\Phi_2} &= \frac{1}{\sqrt{2}}(\ket{0}\ket{0} - \ket{1}\ket{1}) ,\\
\ket{\Phi_3} &= \frac{1}{\sqrt{2}}(\ket{0}\ket{1} - \ket{1}\ket{0}),\\
\sum_i p_i &= 1 \text{ and } p_i \ge 0.
\end{split} \eeq

Therefore our C-Q box can be written as creating states
\beq \begin{split}
\rho^{x,y} &= \hU^{x,y}_A \hV^{x,y}_B \left( \sum_{i=0}^3 p^{x,y}_i \ket{\Phi_i}\bra{\Phi_i} \right) (\hU^{x,y}_A)^{\dagger} (\hV^{x,y}_B)^{\dagger} \\
&= \sum_{i=0}^3 p^{x,y}_i \ket{\chi^{x,y}_i} \bra{\chi^{x,y}_i}, \text{where} \\
\ket{\chi^{x,y}_i} &= \hU^{x,y}_A \hV^{x,y}_B \ket{\Phi_i} \text{and} \\
\sum_i p^{x,y}_i &= 1.
\end{split} \eeq

To implement this in terms of C-C boxes and pre-shared entanglement first we note that we can produce any C-Q box that for each $(x,y)$ has a single pure state $\ket{\chi^{x,y}_i}$ as output. 
This is because the $\ket{\chi^{x,y}_i}$ are all related to $\Phi_0$ by a unitary applied on $A$, and we showed how to implement such boxes in section \ref{MainIdea}.  

\begin{figure}[ht]
\includegraphics[width=8.6cm]{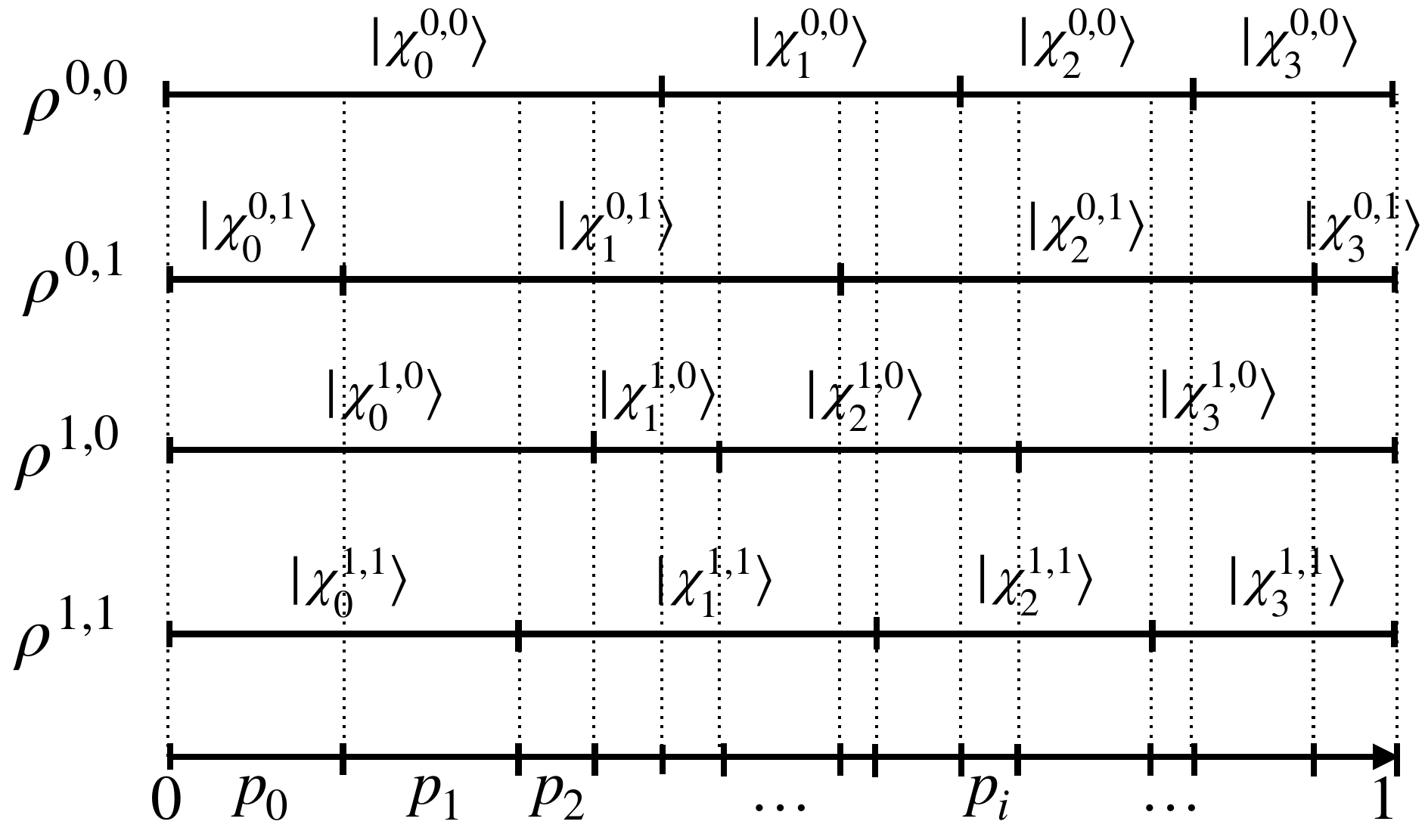}
\caption{Given mixed states $\rho^{0,0}$, $\rho^{0,1}$, $\rho^{1,0}$ and $\rho^{1,1}$, each of which has its probabilistic mixture into pure states $\ket{\chi^{x,y}_i}$ displayed along a horizontal line with the distance between vertical markers representing $p^{x,y}_i$, we show in the bottom line probabilities $p_i$ for each of which all the $\rho^{x,y}$ mixtures have a fixed pure state.}
\label{Fig:CQBoxes3}
\end{figure}

Next we need to probabilistically mix these boxes.  There are many ways to do this.  One way which can be described graphically is shown in Fig. \ref{Fig:CQBoxes3} for $x,y = 0,1$.  This displays the set of mixed states as a mixture of sets of pure states, with mixture probabilities $p_i$.  With probability $p_0$ we create a C-Q box which outputs the states in the first column in the figure, $\ketNoResize{\chi^{0,0}_0}$ for $(x,y) = (0,0)$, $\ketNoResize{\chi^{0,1}_0}$ for $(x,y) = (0,1)$, $\ketNoResize{\chi^{1,0}_0}$ for $(x,y) = (1,0)$, and $\ketNoResize{\chi^{1,1}_0}$ for $(x,y) = (1,1)$. With probability $p_1$ we create a C-Q box which outputs the states from the second column in the figure, which only differ from the first column in the case $(x,y) = (0,1)$, where we create $\ketNoResize{\chi^{0,1}_1}$.  Similarly we read off the states for the other values of $p_i$.

Thus we can create any C-Q box outputting mixed states of 2 maximally disordered qubits.

\section{Multi-Partite States}

C-Q boxes with multi-partite states are more difficult to classify than bi-partite states, due to more parties being involved and the complexity of multi-partite entanglement.  Furthermore the no-signalling conditions lead to many constraints on the sets of states which are allowed.  In particular, we must not only have that there is no-signalling from $A$ to $B$, but also that there is no-signalling from $A$ to the joint system $BC$, and in general from any set of parties to any other set of parties.  
Therefore even before asking whether a multi-partite C-Q box is genuine or not, even constructing the most general C-Q box while making sure it is non-signalling, is an issue in itself. 

Coming now to the question of whether multi-partite C-Q boxes are genuine or not, on one hand the many no-signalling conditions and complexity of entanglement may make simulating them more difficult, and hence allow a genuine C-Q box.  On the other hand, the same conditions limit the number of possible C-Q boxes, which may make them easier to simulate.

In this section we present two examples: W-type and GHZ-type C-Q boxes.  In the first example, C-Q boxes which output W-type quantum states, the constraints are stronger than they are for the bi-partite case, so strong that they limit the class of such possible C-Q boxes in such a way that all the allowed boxes are in some sense weak.  That is,  they are LOSE: they can be implemented using $\bf{L}$ocal $\bf{O}$perations (i.e. unitaries) and pre-$\bf{S}$hared $\bf{E}$ntanglement, where the local operations only depend upon the local inputs and do not require any C-C box that generates correlations beyond quantum mechanics for correlating the local operations.

The second example, GHZ-type C-Q boxes, shows that multi-partite constraints do not always force C-Q boxes to be so weak.  Similar to the bi-partite case, these C-Q boxes require C-C boxes with stronger than quantum mechanical correlations in their simulation.

\subsection{W-Type C-Q Boxes}
For our first example we consider constructing a C-Q box whose outputs are W-type states which differ by arbitrary phases in the computational basis, i.e. 
\beq \begin{split}
\label{WStatePhases}
&\ket{\Psi^{x,y,z}}_{ABC} = \\
&\frac{1}{\sqrt{3}} \left( e^{i \alpha(x,y,z)} \ket{100} + e^{i \beta(x,y,z)} \ket{010} + e^{i \gamma(x,y,z)} \ket{001} \right). 
\end{split} \eeq

This tri-partite ``W-phase'' C-Q box seems similar to the bi-partite phase boxes considered earlier. Yet, while the bi-partite C-Q box allows for correlated phases, we shall show that the only such non-signalling W-phase C-Q boxes are those whose phases are equivalent to local phases: 
\beq
\label{WStateReducedPhases}
\frac{1}{\sqrt{3}} \left( e^{i \alpha(x)} \ket{100} + e^{i \beta(y)} \ket{010} + e^{i \gamma(z)} \ket{001} \right). 
\eeq

To prove this we will create the most general no-signalling box based on $\ket{W}$ with phases by using a no-signalling argument line by line in the following table.

\medskip

$\begin{array}{c || c @{} c c c @{} c c c @{} c} 
 \hline
 xyz & & & & & & & \\
 \hline\hline
 000 & & \ket{100} & + & & \ket{010} & + &  & \ket{001} \\ 
 \hline
 001 & & \ket{100} & + & & \ket{010} & + & e^{i\gamma} & \ket{001} \\ 
 \hline
 010 & & \ket{100} & + & e^{i\beta} & \ket{010} & + & & \ket{001} \\ 
 \hline
 100 & e^{i\alpha} & \ket{100} & + & & \ket{010} & + & & \ket{001} \\ 
 \hline
 011 & & \ket{100} & + & e^{i\beta} & \ket{010} & + & e^{i\gamma} & \ket{001} \\ 
 \hline
 101 & e^{i\alpha} & \ket{100} & + & & \ket{010} & + & e^{i\gamma} & \ket{001} \\ 
 \hline
 110 & e^{i\alpha} & \ket{100} & + & e^{i\beta} & \ket{010} & + & & \ket{001} \\ 
 \hline
 111 & e^{i\alpha} & \ket{100} & + & e^{i\beta} & \ket{010} & + & e^{i\gamma} & \ket{001} \\ 
 \hline
\end{array}$

\medskip

In the first line, $\ket{\Psi^{0,0,0}}_{ABC}$  can be taken without loss of generality to be equal to the $\ket{W}$ state. The second line, $\ket{\Psi^{0,0,1}}_{ABC}$, is the output corresponding to a change of the input of C only.  Hence, when we group A and B together, the reduced density matrix of AB must be the same for $\ket{\Psi^{0,0,0}}_{ABC}$ and $\ket{\Psi^{0,0,1}}_{ABC}$.  The state $\ket{\Psi^{0,0,0}}_{ABC}$ can be written as 
\beq
\ket{\Psi^{0,0,0}}_{ABC}= \frac{1}{\sqrt{3}} (\ket{10}_{AB} + \ket{01}_{AB}) \ket{0}_C + \frac{1}{\sqrt{3}} \ket{00}_{AB} \ket{1}_C .
\eeq
If we change the relative phase between $\ket{100}$ and $\ket{010}$ it will change the density matrix $\rho_{AB}$ and hence be observable.  So by no-signalling the only phase change we can make (up to an overall phase) between $\ket{\Psi^{0,0,0}}_{ABC}$ and $\ket{\Psi^{0,0,1}}_{ABC}$ is on $\ket{001}$.  We call this phase factor $e^{i \gamma}$. Continuing in a similar way, we see that the only non-signalling W-phase box is (up to overall phases) the one in the above table.

Now the phases in the table are all local ($\alpha$ is present exactly when $x=1$, $\beta$ when $y=1$ etc).  So not only are the W-phase C-Q boxes non-genuine, they are LOSE i.e they need no non-local C-C box if we want to create them in a non-genuine way. They can be implemented by local unitaries acting on pre-shared W states.

\subsection{GHZ-Type C-Q Boxes}
A different generalization of the bipartite phase C-Q box applies a phase on the tri-partite GHZ state, i.e.
\beq 
\label{GhzBox} 
\ket{\Psi^{x,y,z}}_{ABC} = \frac{1}{\sqrt{2}} \left(\ket{000} + e^{i \phi(x,y,z)} \ket{111}\right).
\eeq

This box is not LOSE, but can be implemented using a pre-shared GHZ state, a tri-partite C-C box, and a generalization of the algorithm in section \ref{PhaseChange}.  To see how this works, consider first how to implement such a box with $\phi(x,y,z)=\pi x \cdot y \cdot z$.  We use a C-C box with $a,b,c=0,1$ with equal probability, and 
\beq
(a+b+c) \bmod{2} = x \cdot y \cdot z.
\eeq
Alice applies a phase $e^{i \pi}$ to $\ket{1}_A$ when $a$ = 1, and similar for Bob and Charlie, which achieves the desired result.

To apply a more general phase, e.g. $e^{2\pi i x \cdot y \cdot z/n}$, we take a C-C box with $x,y,z=0,1$, $a$ and $b$ chosen independently uniformly from $0..n-1$, and $c$ defined by 
\beq
(a+b+c) \bmod{n} = x \cdot y \cdot z.
\eeq
Alice applies the phase $e^{2\pi i a/n}$ to $A$, and similar for Bob and Charlie.  Note that in order for this C-C box to be valid, it must be non-signalling from any group of parties to any other group.  This is easily checked, for example from $A$ to $BC$ it is non-signalling as the randomness in $a$ hides the value of $x$ in the combined outcomes $b,c$.

Thus we have shown that these GHZ boxes are non-genuine, and, similar to the bipartite phase change box, their implementation requires a C-C box with stronger than quantum correlations.

\section{Approximate C-C + entanglement implementation of C-Q boxes} 
\label{irrationalPhaseChange}

Thus far in the paper we have shown that various C-Q boxes are not genuine, since we were able to implement them via a C-C box and pre-shared entanglement. Here we introduce a new idea, the {\bf approximate} C-C + entanglement implementation of a C-Q box. This is an important idea because it is likely that there are many C-Q boxes that cannot be implemented {\it exactly} by any C-C box with a finite number of outputs + pre-shared entanglement, but which can be implemented arbitrarily closely.  Indeed, even if one can implement a particular C-Q box exactly with one set of resources, it may be possible to implement it as closely as desired with a significantly smaller set of resources.  

An example is the phase change box from subsection \ref{PhaseChange}.  In that section the phase $\theta$ was a rational number.  However suppose the phase was irrational.   It is clear that we cannot implement the C-Q box via a C-C box with a finite number of different discrete outputs.  However, we can approximate any irrational $\theta$ arbitrarily closely by a rational number, and use the procedure described in subsection \ref{PhaseChange} to exactly implement this rational phase box. 

An alternative method which implements the phase change for irrational $\theta$ exactly is to use a C-C box with $a$ and $b$ real numbers in $[0,1)$ satisfying $(a-b) \bmod{1} = (x \cdot y) \theta $.  This is similar to going to the limit of using a rational approximation and letting $n \rightarrow \infty$.

For a more general C-Q box, one can also consider an approximate version of the general method for pure states described in section \ref{generalPureStateTheorem}. For any finite set of inputs $x,y$ we conjecture that it is possible to approximate the desired outputs arbitrarily well using a C-C box with finite dimensional outputs $a,b$, essentially by taking a representative sample of all unitaries distributed according to the Haar measure and pairing them up in a way which is reasonably close to the exact continuous solution.  If one wishes to have a set of unitaries which makes a representative sample of the Harr measure, in the sense that e.g. the average over the sample of any polynomial of degree $n$ matches the average over the Haar measure, then one can use a quantum t-design \cite{quantumTDesign}.
However it's not immediately clear how to generalize that to give us samples over two Harr measures, one for Alice and one for Bob, for each input $x$ and $y$, which are correlated in a particular way and which obey the no-signalling constraints.  We believe it would be quite useful to have such a protocol in general. 

\section{Conclusion}
We have considered the issue of non-locality beyond quantum mechanics, and have introduced the concept of classical input - quantum output (C-Q) non-signalling boxes.  We have investigated the question of whether such boxes are genuine new objects, or all of them could be constructed by objects already known, such as non-local non-signalling classical input - classical output (C-C) boxes and pre-shared quantum entangled states. We have showed that a large class of C-Q boxes, including all bi-partite boxes outputting pure states with arbitrary dimensional inputs, are non-genuine.  

As far as multi-partite C-Q boxes outputting pure states are concerned, we found a subclass of states, generalized GHZ-type phase boxes, which can be simulated with C-C boxes and shared entanglement, similar to the bi-partite ones.  At the same time we found another example, C-Q boxes outputting W-type states, for which the no-signalling constraints are so strong that only a very limited set of such boxes is possible.  Furthermore the boxes in this limited set are somewhat weaker: they are LOSE (can be simulated using Local Operations and Shared Entanglement), and don't need any correlations beyond quantum mechanics.  This may point out limitations in the way in which correlations beyond quantum mechanics may interact with quantum mechanical systems.  We conjecture that the majority of multi-partite C-Q boxes which output pure states are LOSE (can be simulated using Local Operations and Shared Entanglement, and don't need any correlations beyond quantum mechanics). 

There are a few more questions that follow immediately. 

The first question concerns the use of resources when C-Q boxes can be be constructed from a C-C box and unitaries acting on pre-shared entanglement. We have shown that even when the C-Q box seems relatively simple, the C-C box needed for its simulation has to have a significant amount of non-locality. Moreover, even small changes to the C-Q box could result in major changes in the non-locality of the C-C needed (see the phase change box considered in sections III-C and III-D). One could presume that this is simply due to the fact that the output of the C-Q box, which is a quantum state, is in some sense ``analog" (allows for phases that are given by real numbers), while the C-C box has a discrete number of outcomes, so more outcomes are necessary for simulating the continuous parameters in the definition of the quantum states. The problem, however, is not so simple, since in addition to the C-C box there are also the unitary transformations that are applied to the pre-shared quantum state, and unitaries are analog objects. Yet they are not enough. We found this behaviour in a particular case, but expect it to be generic.

Second, and most important: What is the class of non-signalling C-Q boxes? We have encountered this problem  when we attempted to decide whether {\it all} C-Q tri-partite boxes with pure-states outputs are are genuine or not. But how can we know that we considered {\it all} possible such boxes?  If we call the set of output states of a non-signalling C-Q box a ``non-signalling" set of quantum states, how can we find all such sets?  That is, the nonsignaling condition in Eq. \eqref{noSignallingConditions} is very clear, and if we are given a set of states we can easily check if they fulfil it.  But how to construct such a general set? What is its general structure? Crucially, the condition does not refer to the structure of each of the individual states separately but on the set as a whole. When dealing with multi-partite mixed states the problem is likely to be very difficult. 

Incidentally, we also note that there are a few other, and quite important, examples of sets of states whose nonlocal properties are defined globally. For example, a set of orthogonal direct product states that cannot be reliably identified by local measurements and classical communication \cite{quantumNonlocalityWithoutEntanglement}.  Another is that of states of two non-identical spin-1/2 particles used to indicate a direction in a 3D space. When dealing with a single direction, a state in which the two spins are parallel and pointing in the desired direction, is as good of indicating that direction as a state in which the spins are anti-parallel, with the first pointing in the desired direction and the second pointing opposite. But if we want to indicate many different directions, the set of anti-parallel spins is better than the set of parallel spins \cite{anti-parallel}. This type of problem has received relatively little attention; we believe this to be a very important general problem for understanding the structure of quantum mechanics.

Thus far, all the C-Q boxes we analysed turned out to be non-genuine, in the sense that they could be constructed out of C-C boxes and pre-shared entangled quantum states. It is possible that ultimately we find that all C-Q boxes can be simulated this way. However, in case there exist C-Q boxes that cannot be simulated by C-C + entanglement, then a new question emerges: Is there a way of simulating some of those C-Q boxes by some other, as yet unspecified, model that is stronger than C-C + entanglement but weaker than the most general C-Q box.

We believe the above questions are just the tip of an iceberg, and that considering C-Q boxes will lead to further insights into the issue of non-locality.

\medskip

\section{Acknowledgements}
We thank Tony Short and Paul Skrzypczyk for helpful discussions.  Sandu Popescu, Daniel Collins and Carolina Moreira Ferrera acknowledge support of the ERC Advanced Grant FLQuant.  \newline

\textit{Note Added} After completing this work we became aware of three related works we would like to mention.  Firstly, D. Beckman, D. Gottesman, M. A. Nielsen, and J. Preskill \cite{causalLocalizableOperations} considered the simple non-local C-Q Box we discuss in section \ref{simpleNonGenuineCQBox} in their section VI B, in the context of looking at quantum to quantum channels.  Secondly 
D. Schmid, H. Du, M. Mudassar, G. Coulter-de Wit, D. Rosset, and M. J. Hoban \cite{postQuantumChannels} (See also \cite{resourceNonclassicalChannels}) discuss C-Q boxes as a part of classifying all non-signalling bi-partite boxes which have one of (classical, quantum, or empty) for each input/output on each side (e.g. classical input and classical output on side A, empty input and quantum output on side B).  They also raised the question whether "genuine" C-Q boxes exist (their "open Question 2"), and showed that the C-Q box of section \ref{simpleNonGenuineCQBox} is equivalent to a PR box.

\bibliographystyle{quantum}
\bibliography{CQBoxes}

\appendix

\section{State invariance under $\hU_A \hU_B^*$}
\label{appendixStateInvariance}
Here we prove that 
\beq 
\hU_A \hU^*_B \frac{1}{\sqrt{n}} \sum_i \ket{i}_A \ket{i}_B = \frac{1}{\sqrt{n}} \sum_i \ket{i}_A \ket{i}_B, 
\eeq
where $n$ is the number of states $i$.  Below $U_{i,j}$ is the $i,j$ matrix element of $U$, and we have dropped the normalization constant $\frac{1}{\sqrt{n}}$.

\beq
\begin{split}
\hU_A \hU^*_B \sum_i \ket{i}_A \ket{i}_B
&= \sum_i \hU_A \ket{i}_A \hU^*_B \ket{i}_B \\
&= \sum_i( \sum_j \hU_{ji} \ket{j}_A \sum_k \hU^*_{ki} \ket{k}_B) \\
&= \sum_{jk}( \sum_i \hU_{ji} \hU^*_{ki} ) \ket{j}_A \ket{k}_B \\
&= \sum_{jk}( \hU \hU^{\dagger} )_{jk} \ket{j}_A \ket{k}_B \\
&= \sum_{jk} \mathbf{1}_{jk} \ket{j}_A \ket{k}_B \\
&= \sum_k \ket{k}_A \ket{k}_B,
\end{split}
\eeq
as desired.

\section{Resources for the phase change box}
\label{appendixResourcesProof}
In section \ref{phaseChangeResources} we claimed that we need a C-C box with $n$ outcomes along with pre-shared entanglement in order to simulate the phase change C-Q box
\beq
\label{phaseChangeBoxDefn}
\ket{\Psi^{x,y}}=\alpha\ket{0}\ket{0}+\beta e^{i 2 \pi x\cdot y / n } \ket{1}\ket{1}. 
\eeq
We proved this for the case where we take a C-C box and the pre-shared entangled state $\alpha\ket{0}\ket{0}+\beta \ket{1}\ket{1}$, and apply unitaries acting only on the state.  However we may wonder if starting with a general state with more dimensions for Alice and Bob, and applying unitaries acting across all the dimensions, allows us to simulate the C-Q box with a simpler C-C box.  Here we shall prove that it will not help: we require the C-C box with $n$ outcomes for $a$ and $b$, defined as
\beq
\label{ccBoxModN}
(a-b) \bmod{n} = x \cdot y.
\eeq

To begin, we take Alice and Bob to share an entangled quantum state
\beq
    \ket{\psi}=(\alpha\ket{00}_{AB}+\beta\ket{11}_{AB})\ket{\phi}_{A'B'}
\eeq
for some arbitrary state $\ket{\phi}$, where Alice has particles $A$ \& $A'$ and Bob has particles $B$ \& $B'$.  This is general, since the addition of $(\alpha\ket{00}_{AB}+\beta\ket{11}_{AB})$ to the arbitrary state $\ket{\phi}$ cannot increase the required C-C box resources.  They also share a C-C box with inputs $(x,y)=(0,1)$ and outputs $a,b=0..(m-1)$ for some $m$.  

For $(x,y)=(0,0)$ we set the first pair of outcomes of the C-C box as $(a,b)=(0,0)$ and let Alice and Bob apply operators $\hU^0$ and $\hV^0$ which implement the desired C-Q box on particles $A$ and $B$, giving 
\beq
\label{zeroZeroSim}
    \hU^0\hV^0\ket{\psi}=(\alpha\ket{00}_{AB}+\beta\ket{11}_{AB})\ket{\phi^{00}}_{A'B'},
\eeq
for some $\ket{\phi^{00}}_{A'B'}$.  Note that particles $A'$ and $B'$ can end up in any state: we only care that we implement the C-Q box in particles $A$ and $B$.

Without loss of generality we can take
\beq
\hU^0 = \hV^0 = \mathbf{1}
\eeq
since if they were anything else we could remove them by the change of basis 
\beq
\begin{split}
\hU^a & \rightarrow \hU^a \hU^{0 \dagger} \\
\hV^b & \rightarrow \hV^b \hV^{0 \dagger} \\
\ket{\psi} & \rightarrow \hU^0 \hV^0 \ket{\psi} \\
&= (\alpha\ket{00}_{AB}+\beta\ket{11}_{AB})\ket{\phi^{00}}_{A'B'}.
\end{split}
\eeq

Next we shall show that for $(x,y)=(0,0)$ the C-C box may as well always give equal outputs for Alice and Bob, $(a,b)=(a,a)$.  Suppose instead that our C-C box gave any other pair of outputs $(a,b)=(0,b)$.  Then Alice and Bob would apply $\hU^0$ and $\hV^b$ on the initial state, giving
\beq
    \hU^0\hV^b\ket{\psi}=(\alpha\ket{00}_{AB}+\beta\ket{11}_{AB})\ket{\phi^{0b}}_{A'B'}.
\eeq
We can substitute for $\ket{\psi}$ using Eq. \eqref{zeroZeroSim} giving
\beq
\begin{split}
\hU^0\hV^b\ket{\psi}
&=\hU^0\hV^b\hU^{0\dagger}\hV^{0\dagger}(\alpha\ket{00}_{AB}+\beta\ket{11}_{AB})\ket{\phi^{00}}_{A'B'}\\
&=\hV^b\hV^{0\dagger}(\alpha\ket{00}_{AB}+\beta\ket{11}_{AB})\ket{\phi^{00}}_{A'B'}\\
&=(\alpha\ket{00}_{AB}+\beta\ket{11}_{AB})\ket{\phi^{0b}}_{A'B'}.
\end{split}
\eeq
From the last two lines we see that $\hV^b\hV^{0\dagger}$, which acts only upon $BB'$, must be of the following form 
\beq
\hV^b\hV^{0\dagger}=\ket{0}_B\braL{B}{0}\hW^b_{B'}+\ket{1}_B\braL{B}{1}\hR^b_{B'},
\eeq
where
\beq
\hW^b_{B'}\ket{\phi^{00}}_{A'B'}=\hR^b_{B'}\ket{\phi^{00}}_{A'B'}=\ket{\phi^{0b}}_{A'B'}.
\eeq
Rewriting the state $\ket{\phi^{00}}_{A'B'}$ in the Schmidt decomposition we have
\beq
\hW^b_{B'}\sum_j\mu_j\ket{j}_{A'}\ket{j}_{B'}=\hR^b_{B'}\sum_j\mu_j\ket{j}_{A'}\ket{j}_{B'},
\eeq
and see that the actions of $\hW^b_{B'}$ and $\hR^b_{B'}$ have to be the same on any $\ket{j}$ such that $\mu_j\neq0$, allowing us to conclude
\beq
    \hV^b=\left(\mathbf{1}_B\otimes \hW^b_{B'}\right)\hV^0.
\eeq
Hence, any other pair of outputs $(0,b)$ has the same action on the output state $\alpha\ket{00}_{AB}+\beta\ket{11}_{AB}$ as $(0,0)$ so it is redundant to include any more pairs than those of the form $(a,a)$ for the case where $(x,y)=(0,0)$.  

A similar argument applies for the $(x,y)=(0,1)$ and $(x,y)=(1,0)$ cases, which must also have pairs of the form $(a,a)$.  For the $(x,y)=(1,1)$ case, comparing with the $(x,y)=(1,0)$ case, no signalling from Bob to Alice means that Alice must get the outcomes $0..(m-1)$.  By symmetry Bob must also get the outcomes $0..(m-1)$.  When $a=0$, we can always relabel the outcomes so that $b=1$.  Then for $a=1$, we can either set $b=0$, or by relabelling set $b=2$. If we choose the latter, we have the choice for $a=2$ to set $b=0$ or $b=3$.  Whatever we do, eventually we need to choose $b=0$, and then the box will have, at least for the first few columns, the following form for some $k$:
\begin{table}[ht]
\label{nOutputBox}
\begin{center}
\begin{tabular}{ |c|c c c l| }
 \hline
 x y & & & (a,b) & \\ 
 \hline
 0 0 & 00 & 11 & \ldots & (k-1)(k-1) \\ 
 0 1 & 00 & 11 & \ldots & (k-1)(k-1) \\
 1 0 & 00 & 11 & \ldots & (k-1)(k-1) \\
 1 1 & 01 & 12 & \ldots & (k-1)0 \\
 \hline
\end{tabular}
\end{center}
\end{table} 

There may be more columns, for example after $k$ columns the box could add another $k$ columns copied from the first $k$ with all entries larger by $k$, but at minimum the first $k$ columns give the same outputs as the C-C box described in Eq. \eqref{ccBoxModN}, with $k$ outcomes instead of $n$.  So we have proved we need a C-C box containing that form, and now need to show that the lowest possible value of $k$ such that Alice and Bob achieve the desired quantum state is in fact $n$.  

We can write the C-Q box defined in Eq. \eqref{phaseChangeBoxDefn} as
\beq
\ket{\Psi^{x,y}} = \hSigma_z^{2 x \cdot y/n}(\alpha\ket{00}_{AB}+\beta\ket{11}_{AB}),
\eeq
where $\hSigma_z$ is the Pauli matrix which does 
\beq \begin{split}
\ket{0} &\rightarrow \ket{0} \\
\ket{1} &\rightarrow e^{i \pi} \ket{1}.
\end{split} \eeq

In order to simulate this we take the C-C box above.  Without loss of generality, for a fixed outcome $a$, Alice might apply different unitaries depending upon the input $x$, i.e. apply $\hU^a_x$.  Similarly Bob could apply $\hV^b_y$.  However, if we compare the cases $x,y = 1,0$ and $x,y=0,0$ for any fixed outcome $a=b$, we get
\beq
\begin{split}
\hU^a_0\hV^a_0\ket{\psi} &= (\alpha\ket{00}+\beta\ket{11})\ket{\phi^{aa}_{00}} \\
\hU^a_1\hV^a_0\ket{\psi} &= (\alpha\ket{00}+\beta\ket{11})\ket{\phi^{aa}_{10}} \\
&= \hU^a_1 \hV^a_0 \hV^{a \dagger}_0 \hU^{a \dagger}_0 (\alpha\ket{00}+\beta\ket{11})\ket{\phi^{aa}_{10}}.
\end{split}
\eeq
Using arguments similar to before we see that Alice's unitaries for $x=0$ and $x=1$ are related by a unitary acting only on $A'$, i.e.
\beq
\hU^a_1 = \left( \mathbf{1}_A \otimes \hT^a_{A'} \right) \hU^a_0,
\eeq
for some $\hT^a_{A'}$.  Since $\hT^a_{A'}$ only affects $A'$ after we are finished creating our desired state in $A$, we can omit it, and write $\hU^a_1 = \hU^a_0 = \hU^a$, dropping the subscripts.  Similarly for Bob we can write $\hV^b_y = \hV^b$.  That simplifies the next step, relating the various $\hU^a$ to one another.

In the case where $x\cdot y=0$, Alice and Bob obtain the following output state
\beq
\label{aASim}
    \hU^a\hV^a\ket{\psi}=(\alpha\ket{00}_{AB}+\beta\ket{11}_{AB})\ket{\phi^{a\;a}}_{A'B'}.
\eeq
In the case where $x\cdot y=1$, they instead find
\beq
    \hU^a\hV^{a+1}\ket{\psi}=\hSigma_z^{2/n}(\alpha\ket{00}_{AB}+\beta\ket{11}_{AB})\ket{\phi^{a\;a+1}}_{A'B'},
\eeq
where the superscript $a+1$ is taken $\bmod{k}$.  We can move $\hSigma_z^{2/n}$ in this equation to the other side, giving
\beq
    (\hSigma_z^{2/n})^{\dagger}\hU^a\hV^{a+1}\ket{\psi}=(\alpha\ket{00}_{AB}+\beta\ket{11}_{AB})\ket{\phi^{a\;a+1}}_{A'B'},
\eeq
and then use Eq. \eqref{aASim} to substitute for $\ket{\psi}$ giving
\beq
\begin{split}
    &(\hSigma_z^{2/n})^{\dagger} \hU^a\hV^{a+1}\ket{\psi}\\
    &=(\hSigma_z^{2/n})^{\dagger} \hU^a\hV^{a+1}\hU^{a\dagger}\hV^{a\dagger}(\alpha\ket{00}_{AB}+\beta\ket{11}_{AB})\ket{\phi^{a\;a}}_{A'B'}\\
    &=(\hSigma_z^{2/n})^{\dagger} \hV^{a+1}\hV^{a\dagger}(\alpha\ket{00}_{AB}+\beta\ket{11}_{AB})\ket{\phi^{a\;a}}_{A'B'}\\
    &=(\alpha\ket{00}_{AB}+\beta\ket{11}_{AB})\ket{\phi^{a\;a+1}}_{A'B'}.
\end{split}
\eeq
From the last two lines, following the same reasoning as before, we must have
\beq
    \hV^{a+1}=\left( \hSigma_z^{2/n}\otimes \hW^{a+1}_{B'}\right)\hV^a.
\eeq

By recursion and the fact that $\hV^0 = \mathbf{1}$ we obtain the following expression for $\hV^b$
\beq
    \hV^b=\hSigma_z^{2b/n}\otimes \hQ^b_{B'},
\eeq
for some $\hQ^b_{B'}$.  Since we do not care about the final state of $B'$, we will drop $\hQ^b_{B'}$ and say 
\beq
\hV^b= \hSigma_z^{2b/n}.
\eeq

By a similar argument we have
\beq
    \hU^a=\hSigma_z^{-2a/n}.
\eeq

We can check that this simulates the C-Q box for $x \cdot y = 0$ as follows:
\beq
\begin{split}
\hU^a \hV^a \ket{\psi} &= \hSigma_z^{-2a/n} \hSigma_z^{2a/n} \ket{\psi} \\
&= (\alpha\ket{00}_{AB}+\beta\ket{11}_{AB})\ket{\phi^{00}}_{A'B'}.
\end{split}
\eeq

For $x \cdot y = 1$ our C-C box has paired up $a$ and $b$ as $(a,(a+1) \bmod{k})$.  This gives us
\beq
\begin{split}
\hU^a \hV^{(a+1) \bmod k} \ket{\psi} &= \hSigma_z^{-2a/n} \hSigma_z^{2((a+1) \bmod k)/n} \ket{\psi} \\
&=  \hSigma_z^{2 (((a+1) \bmod k) - a)/n} \ket{\psi},
\end{split}
\eeq
for $a=0..(k-1)$.  The exponent is
\beq
 2 \frac{((a+1) \bmod k) - a}{n} = \begin{cases} 
 2/n & \text{when } a + 1<k,\\
 2/n - 2k/n & \text{when } a + 1=k.
 \end{cases}
\eeq
Since $\hSigma_z^2 = 1$, this gives the desired state,
\beq
\hSigma_z^{2/n} (\alpha\ket{00}_{AB}+\beta\ket{11}_{AB})\ket{\phi^{00}}_{A'B'},
\eeq
so long as $k$ is a multiple of $n$.  Thus the smallest C-C box which can implement our C-Q box has $n$ outputs, completing our proof.

\end{document}